\documentclass[11pt]{article}
\date{ }
\author{Vladimir E. Zakharov$^{1, 2}$}
\title{Direct and Inverse Cascades in the Wind-Driven Sea}

\begin{document}

\maketitle

\noindent \small{ $^1$ Department of Mathematics, University of
Arizona, Tucson, AZ 85721, USA, {\it
zakharov@math.arizona.edu};\\
$^2$ Landau Institute for Theoretical Physics, Moscow, Russia}

\begin{abstract}

 We offer a new form for the $S_{nl}$ term in the Hasselmann kinetic
equation for squared wave amplitudes of wind-driven gravity wave. This
form of $S_{nl}$ makes possible to rewrite in differential form the
conservation laws for energy,
momentum, and wave action, and introduce their fluxes by a natural way. We
show that the stationary kinetic equation has a family of exact
Kolmogorov-type solutions governed by the fluxes of motion constants: wave action, energy, and momentum.
 The
simple "local" model for $S_{nl}$ term that is equivalent to the
"diffusion approximation" is studied in details. In this case, Kolmogorov
spectra are found in the explicit form. We show that a general solution of
the stationary kinetic equation behind the spectral peak is described by
the Kolmogorov-type solution with frequency-dependent fluxes. The domains
of "inverse cascade" and "direct cascade" can be separated by a natural
way. The spectrum in the universal domain is close to $\omega^{-4}$.

\end{abstract}


\section{INTRODUCTION}

    The phenomenon of wind-generated gravity waves on the sea surface
is a very interesting subject not only for oceanographers, coastal
engineers and naval architects. The ocean is a great natural laboratory,
and many phenomena taking place there are interesting for a broad
community of physicists. Gravity and capillary surface waves on
deep water represent the most conspicuous natural example of
nonlinear waves in a strongly dispersive media. The statistical theory of
such waves is called a theory of weak turbulence; it is an important part of
general classical physics. This theory has been developing for more
than forty years, and its basic concepts are now understood very well.
However experimental data supporting this theory are scarce. The situation
here is opposite to that of strong hydrodynamic turbulence in an
incompressible fluid. In this case we have a lot of experimental data
 but the theory is poor and inconsistent.

    The theory of weak turbulence is naturally applicable to the
description of surface gravity and capillary waves on deep and shallow
water. It can also be applied to acoustic Alfven waves in hydrodynamics,
to many types of waves in plasmas, to waves in liquid helium and spin
waves in ferromagnetics, as well as to Rossby waves in atmosphere and to
internal waves; however experimental data collected so far in these areas of physics
is too poor to perform a convincing comparison with the theory. Only
the oceanographers who collect data on wave spectra, measured in the ocean, in lakes
and in wave tanks for almost half a century, have accumulated vast quantities
of very valuable experimental materials which could and should be compared
with the predictions of the theory. (A contribution of professor M. Donelan
in this process is really seminal, see for instance [Donelan, 1985]) This is an ambitious task; a lot
of work must be done to perform it, and only first results are obtained in this direction.
It was shown that the fetch
dependence of wave energy and wave frequency, obtained in basic
fetch-limited experiments, can be naturally interpreted in terms of
the weak turbulent theory [Zakharov, 2002]. All experimenters agree that the observed
spectra of wind-driven sea waves just behind the spectral peak have
a universal powerlike form close to $\omega^{-4}$. We will show that this dependence
can be easily explained in terms of the weak-turbulent theory.

    Once more, weak turbulence is the statistical theory of nonlinear
waves in dispersive media. A central point of this theory is the following:
a wave ensemble is described by the kinetic equation for square wave
amplitudes. This equation has different names; for instance, the Boltzmann
equation or the Hasselmann equation. Also, this equation could be called the Peierls
equation, because it is nothing but the classical limit of the quantum
kinetic equation for phonons, derived by Peierls and others in
the late twenties. In this article we will use the term "kinetic
equation".

    The kinetic equation for gravity waves was derived by K. Hasselmann in
1962-1963 [Hasselmann, 1962; Hasselmann 1963]; now this equation is accepted as a basic
model for the description of wave spectra evolution by the majority of oceanographers.
The kinetic equation in a
truncated form (known as the DIA, or Direct Interaction Approximation) is widely
used in the third generation models of wave prediction. However the
physical effects described by the kinetic equation need to be commented.

    The main function of the nonlinear interaction term $S_{nl}$ in the kinetic
equation is a very intensive redistributor of energy, momentum and
wave action along the spectrum. Due to $S_{nl}$, the direct cascades of energy and
momentum as well as inverse cascades of wave action are formed. These
processes govern the evolution of the spectral peak and play a central
role in the formation of the universal powerlike spectrum behind the
spectral peak. In the simplest idealized cases, these spectra are
Kolmogorov-type weak turbulent spectra, corresponding to constant fluxes
of energy and wave action. Strictly speaking, such spectra are realized
in physical systems, where domains of forcing and damping are essentially
separated in $K$-space. In the wind-driven sea the source of wave action is concentrated
near the spectral peak, while the source of energy is distributed along the spectrum.
On the first glance this fact is an impediment for application of Kolmogorov-type theory;
actually this is not a serious difficulty. In a realistic situation the fluxes of energy and
 wave action are functions of
frequency. However this does not affect essentially the spectral shape; it remains close to $\omega^{-4}$.

    One more point is important. If the wind-driven sea is
well-developed, then the main part of momentum fluxes from the wind is concentrated
in short waves. This fact, experimentally established  and essentially stressed by
M. Donelan, can be naturally explained in terms of the weak turbulent theory.

    There is no reliable parametrization for the white capping dissipation. However it
seems very probable that this fundamental process generates
dissipation only in the short-wave region. Indeed, wave breaking makes the
surface more smooth, acting like an efficient viscosity or even
super-viscosity. Moreover, the wave breaking generates turbulence in the
boundary layer with a thickness comparable to the length of the most
dissipative waves. It is known that this layer is much thinner than the
wave length of the spectral peak. Hence it is possible to suggest that
 the white cap dissipation in the area of the
spectral peak can be neglected for the developed sea. In the first approximation the evolution of
the spectral peak is described by the "conservative" kinetic equation such
that includes only the time derivative, the advective term, and $S_{nl}$. In the range of high frequencies the influence
of wind forcing and white capping can be taken into account as a "boundary
condition". This condition defines the flux of wave
action to the long wave region.

    The "conservative" kinetic equation has a family of self-similar
solutions depending on two free parameters. It was shown recently that by choosing
the parameters in a proper way one can explain the results of
major fetch-limited experiments made during the last three decades,
including the JONSWAP experiment [Zakharov, 2002]. We will present a detailed description of this study,
supported by a massive numerical experiment, in an another article. In this paper we present
the basic ideas
of the weak turbulent theory using the simplest theoretical model of $S_{nl}$ known as
"local" or "diffusive" approximation.

\section{Basic Theory}

    We assume that the flow in the wave motion is potential $v=\nabla \Phi$. The
condition of incompressibility imposes on potential $\Phi$ the
Laplace equation:

\begin{equation}
 \Delta \Phi=0.
\end{equation}
If $\eta$ is the shape of the surface, then equation (1) should be solved in the
domain $z<\eta$ under the boundary condition

\begin{eqnarray}
&&\Phi \vert_{z=\eta}=\psi(\vec r, t),\nonumber\\
&&\left.\frac{\partial \Phi}{\partial z} \right |_{z \to - \infty}
\rightarrow 0 .\nonumber
\end{eqnarray}
 In the linear approximation we should solve equation (1) in the
half-space $z<0$. The shape of the surface and the potential on the
surface, $\eta$ and $\psi$, are canonically conjugated variables; then the Euler equation for
the potential flow of an ideal fluid with a free surface can be written

\begin{eqnarray}
&&\eta_{t}=\frac{\delta H}{\delta \psi},\nonumber\\
&&\psi_{t}=-\frac{\delta H}{\delta \eta}.
\end{eqnarray}

    The solution of linearized motion equation (2) is the propagating
monochromatic wave

\begin{eqnarray}
&&\eta = \sqrt{\frac{2 \omega_{\vec k}}{g}}\,A_{0}\cos(\vec k \vec r-\omega_{k}t-\phi),\nonumber\\
&&\psi = \sqrt{\frac{2 \omega_{\vec k}}{|\vec k|}}\,A_{0}\sin(\vec k \vec
r-\omega_{k}t-\phi),\nonumber
\end{eqnarray}
 where $\omega_{\vec k}=\sqrt{g|\vec k|}$ and $\vec k$ is the wave vector. We can call
$$
A=A_{0}\,e^{i\phi}
$$
the complex amplitude of the propagating wave. The normalization of $A$ is
taken in such way that the energy density is
$$
E=\omega A_0 ^{2}.
$$
By definition, $A_0 ^{2}=E/\omega$ is the density of "wave
action"; then the sea surface is a composition of propagating waves

\begin{eqnarray}
&&\eta=\int\sqrt{\frac{\omega_{\vec k}}{2g}}\,(A_{\vec k}+A^{*}_{-\vec k})\,
e^{i(\vec k \vec r-\omega_{\vec k}t)}\, d\vec k ,\nonumber\\
&&\psi=-i\int\sqrt{\frac{\omega_{\vec k}}{2|\vec k|}}\,(A_{\vec k} -
A^{*}_{-\vec k})\, e^{i(\vec k \vec r - \omega _{\vec k} t)}\, d\vec k
.\nonumber
\end{eqnarray}

A real sea should be described statistically; to do this let us
introduce the spectral density of wave action, assuming that
$$
\left\langle A_{\vec k}\, A^{*}_{\vec k^{'}} \right\rangle = g\, N_{\vec
k}\,\,\,\delta(\vec k-\vec k^{'}).
$$
Then we can express the spatial correlation function
$$
F(\vec R)= \left \langle \eta(\vec r)\, \eta(\vec r+\vec R) \right \rangle
$$
in the form
$$
F(\vec R)=\int \omega_{\vec k}\,\, N_{\vec k}\, \cos \vec k \vec R\, d\vec k  .
$$
In this case,
the mean squared derivation $\sigma$ is given by the formula
$$
\sigma=\left\langle \eta^{2}\right\rangle = \int\omega_{\vec k}\, N_{\vec
k}\,
d\vec k .
$$

Further, let us denote $E_{\vec k}=\omega_{\vec k}\, N_{\vec k}$; this is the energy density in
$K$-space divided by $g$. It has dimension $\left[L^{4}\right]$.
Now we can express $\eta(\vec r)$ through its Fourier transform
$\eta_{\vec k}$
$$
\eta(\vec r)=\int\eta_{\vec k}\,\,e^{i\vec k \vec r}\,d\vec k
$$
and define the spatial spectrum as follows
\begin{eqnarray}
&&\left\langle\eta_{\vec k}\,\eta_{\vec k^{'}}\right\rangle = I_{\vec
k}\,\,\delta(\vec k+\vec k^{'}),\nonumber\\
&&F(\vec R)=\int I_{\vec k}\,\, e^{-i \vec k \vec r}\, d\vec k .\nonumber
\end{eqnarray}
Comparing with (3) we obtain
\begin{eqnarray}
&&I_{\vec k}=\frac{1}{2}\,\omega_{\vec k}\,(N_{\vec k}+N_{-\vec k}),\nonumber\\
&&\sigma=\int I_{\vec k}\,\,d\vec k .
\end{eqnarray}

Further, it is convenient to introduce complex amplitudes
$$
a_{\vec k}=2\pi\, A_{\vec k}
$$
and derive the motion equation (2) in the form
\begin{equation}
\frac{\partial a_{\vec k}}{\partial t}+i\, \frac{\delta H}{\delta
a^{*}_{\vec k}} =0,
\end{equation}
where the Hamiltonian $H$ can be expanded in powers of $a_{\vec k}$
\begin{eqnarray}
H&=&H_{0}+H_{1}+H_{2}+\dots ,\nonumber\\
H_{0}&=&\int \omega_{\vec k} \left|a_{\vec k}\right|^2 \, d\vec k ,\nonumber\\
H_{1}&=&\frac{1}{2}\int V_{\vec k \vec k_{1} \vec k_{2}}\left(a^{*}_{\vec
k} a_{\vec k_{1}} a_{\vec k_{2}}+ a_{\vec k} a^*_{\vec k_{1}} a^*_{\vec k_{2}}
 \right)\nonumber\\
&& \hspace{1in}\times \delta(\vec k - \vec k_{1} - \vec
k_{2})\,d\vec k\, d\vec k_{1}\, d\vec k_{2}\nonumber\\
&+&\frac{1}{3}\int U_{\vec k \vec k_{1} \vec k_{2}} \left(a_{\vec k}\,
a_{\vec k^{1}}\,a_{\vec k^{2}}+a^{*}_{\vec k}\,a^{*}_{\vec k_{1}}\,a^{*}_{\vec
k_{2}}\right)\nonumber\\
&&\hspace{1in}\times\delta(\vec k+\vec k_{1}+\vec k_{2})\,d\vec k\, d\vec k_{1}\,d\vec
k_{2} .\nonumber
\end{eqnarray}
The Hamiltonian $H_{2}$ contains terms quartic in $a^{*}_{\vec k}$,  $a_{\vec
k}$.

   It is not very convenient to use equation (4). The cubic
Hamiltonian $H_{1}$ leads to the formation of "slave" waves; wave
numbers and frequencies of "slave waves" are not connected by the dispersion relation.
To separate "slave" and "free" waves one should perform a canonical
transformation to new variables $b_{\vec k}$, eliminating the cubic term
$H_{1}$. In new variables the Hamiltonian takes the
form [Zakharov, 1999]:
 \begin{eqnarray}
H&=&H_{0}+H_{2} ,\nonumber\\
H_{0}&=&\int \omega_{\vec k}\, b_{\vec k}\, b^{*}_{\vec k}\, d\vec k ,\nonumber\\
H_{2}&=&\frac{1}{4}\int T_{\vec k \vec k_{1} \vec k_{2} \vec k_{3}}\,\,b^{*}_{\vec
k}\, b^{*}_{\vec k_{1}}\, b_{\vec k_{2}}\, b_{\vec k_{3}}\nonumber\\
&&\hspace{0.5in}\times \delta(\vec k + \vec k_{1} - \vec
k_{2}-\vec k_{3})\,d\vec k\, d\vec k_{1}\, d\vec k_{2}\, d\vec k_{3} ,\nonumber
\end{eqnarray}
and the dynamic equation
\begin{equation}
\frac{\partial b_{\vec k}}{\partial t}+i\frac{\delta H}{\delta b^{*}_{\vec
k}}=0
\end{equation}
in new variables is known as  "Zakharov equation" [Zakharov, 1968]
\begin{eqnarray}
&&\frac{\partial b_{\vec k}}{\partial t}+i\,\omega_{\vec k}\, b_{\vec
k}+\frac{i}{2}\int T_{\vec k \vec k_{1} \vec k_{2} \vec k_{3}}\,\,b^{*}_{\vec
k_{1}}\, b_{\vec k_{2}}\, b_{\vec k_{3}}\nonumber\\
&&\hspace{0.5in}\times \delta(\vec k + \vec k_{1} -\vec k_{2}- \vec k_{3})\,d\vec k_{1}\,
d\vec k_{2}\,d\vec k_{3}=0.\nonumber
\end{eqnarray}
Explicit expressions for the coefficients of the Hamiltonian,
the coupling coefficient $T_{\vec k \vec k_{1}
\vec k_{2} \vec k_{3}}$, and the canonical transformation can be found in (Zakharov, 1999).

Equation (5), being approximate, has a very important feature: it
conserves the total wave action, i.e., is adiabatic invariant
$$
N=\int\left|b_{\vec k}\right|^2\,d\vec k  ,
$$
thus the kinetic wave equation is imposed to the correlation function of
$b$-variables:
$$
\left\langle b_{\vec k}\,b^{*}_{\vec k^{'}}\right\rangle = n_{\vec
k}\,\,\delta(\vec k-\vec k^{'}).
$$
On deep water we can put approximately
$$
n_{\vec k} \simeq 4 \pi^{2} g\, N_{\vec k} .
$$

It is important to stress that the kinetic equation describes not the real sea
 studied by experimenters but an idealized object: the
ensemble of "free" waves filtered from the slave harmonics. On deep water
the difference between these two ensembles is small $(1-2\%)$, while on
shallow water the difference can be much more essential.

The kinetic equation in terms of $N$ reads
$$
\frac{\partial N}{\partial t} + \frac{\partial \omega}{\delta \vec k} +
\nabla N = S_{nl}+S_{in}+S_{ds}.
$$
Here $S_{in}$ and $S_{ds}$ are income from wind and dissipation due to
white capping, and $S_{nl}$ has the form
\begin{eqnarray}
S_{nl}&=&\int S\left(\vec k, \vec k_{1}, \vec k_{2}, \vec k_{3}\right)
\left(N_{\vec k_{1}} N_{\vec k_{2}} N_{\vec k_{3}} + N_{\vec k}
N_{\vec k_{2}} N_{\vec k_{3}}\right.\nonumber\\
&& -\left. N_{\vec k} N_{\vec k_{1}} N_{\vec k_{2}} -
N_{\vec k} N_{\vec k_{1}} N_{\vec k_{3}}\right)\,\,
\delta\left(\vec k + \vec k_{1}-\vec k_{2} - \vec k_{3}\right)\nonumber\\
&&\hspace{0.5in}\times\delta
\left(\omega _{\vec k}+\omega _{\vec k_{1}}-\omega _{\vec k_{2}} -
\omega _{\vec k_{3}}\right)\, d\vec k_{1}\, d\vec k_{2}\, d\vec k_{3} ,\nonumber
\end{eqnarray}
where
$$
S\left(\vec k, \vec k_{1}, \vec k_{2}, \vec k_{3}\right) = (2\pi)^{4}\pi
g^{2} \left|T_{\vec k \vec k_{1} \vec k_{2} \vec
k_{3}}\right|^{2}
$$
can be found in [Hasselman, 1963], [Webb, 1978], [Zakharov, 1999].

    The explicit expression for $S$ is pretty complicated. The most
important fact is the following: $S\left(\vec k, \vec k_{1}, \vec k_{2}, \vec
k_{3},\right)$ is a  homogeneous function of sixth order
$$
S\left(\epsilon \vec k, \epsilon \vec k_{1}, \epsilon \vec k_{2},
\epsilon \vec k_{3},\right)=\epsilon^{6} S\left(\vec k, \vec k_{1}, \vec k_{2}, \vec
k_{3}\right).
$$
 For a rough estimate we can put
$$
S\simeq k^{6}\simeq \omega^{12}.
$$
 This is very fast growing function in frequencies. This fact is of a
crucial importance.

Most authors agree that $S_{in}$ can be presented in the form
$$
S_{in}=\beta(\omega, \theta)\,N(k),
$$
where
$$
\beta(\omega, \theta)=\mu\,F(\xi)\,\omega,\quad \xi=\frac{\omega
\cos\theta}{\omega_0},\quad \omega_0=\frac{g}{u_{10}}.
$$
Here $u_{10}$ is the wind velocity at 10 meters height; $\mu=0.1\sim
0.3$; $\rho_a/\rho_w \simeq 10^{-3}$; $\rho_a$ and $\rho_w$ are densities of
air and water.

There is no agreement about the exact form of the function $F(\xi)$. According
to Hsiao and Shemdin [Hsiao, 1983]
$$
\mu=0.12,\quad F(\xi)=(0.85\xi -1)^2,
$$
according to Donelan [Donelan, 1985]
$$
\mu\simeq 0.194,\quad F(\xi)=(\xi -1)^2,
$$
while Tolman and Chalikov [Tolman, 1996] proposed a complicated form of $F(\xi)$. In all
these models $\beta(\omega)\simeq \omega^3$ as  $\omega\to\infty$.

According to Snyder [Snyder, 1981]
$$
\mu=0.25,\quad F(\xi)=\xi -1 .
$$
In this case, $\beta(\omega)\simeq \omega^2$ at large $\omega$.

An analytical expression for $S_{ds}$ is much less certain. Komen et al
(1984) proposed the form
\begin{equation}
S_{ds}=-C_{dis}\,\left(\frac{\hat\alpha}{\alpha_{pm}}\right)^4\,\left(\frac{\omega}
{\bar\omega}\right)^n\,\bar\omega\,N.
\end{equation}
Here $\alpha$ is dimensionless steepness  and $C_{dis}$ is a dimensionless
parameter. This formula is entirely speculative. It doesn't have any theoretical
foundation and is not derived from any real experiment in the ocean, lake
or in a wave tank. Anyway, this formula is used widely in operational
models (WAM, SWAM). It is supposed in most cases that
$$
n=2, \quad \alpha=3.33\times 10^{-5},\quad \alpha_{pm}=4.5\times 10^{-3}.
$$
In our opinion, expression (6) overestimates dissipation due to white capping in low frequencies.
It can be used in absence of a better option; however  on our opinion
the parameter $n$ should be essentially larger. If $n\geq 3$, the whole
picture of ocean wave turbulence does not depend on a particular value of
$n$.

\section{Constants of motion and their fluxes}

    In this chapter we study the conservative homogeneous equation
\begin{equation}
\frac{\partial N}{\partial t}=S_{nl}
\end{equation}
in absence of wind forcing and dissipation. It is considered that this
equation has the following constants of motion - wave action, energy and
momentum:
\begin{eqnarray}
&&N=\int N_{\vec k}\,d\vec k ,\nonumber\\
&&E=\int \omega_{\vec k} N_{\vec k}\, d\vec k ,\nonumber\\
&&\vec M=\int \vec k N_{\vec k}\, d\vec k .\nonumber
\end{eqnarray}
Formally speaking, this statement is correct; but the reality is much more complicated.
Let us introduce polar coordinates $|k|$ and $\phi$:
\begin{eqnarray}
&&|k|=\frac{\omega^{2}}{g} ,\nonumber\\
&&k\,dk\,d\phi = \frac{2\omega^{3}}{g^{2}}\,d\omega\, d\phi ,\nonumber
\end{eqnarray}
and denote
\begin{eqnarray}
&&N_{\omega}\,d\omega=N_k \, d\vec k ,\nonumber\\
&&N(\omega, \phi)=\frac{2\omega^{3}}{g}\,N\left(\frac{\omega^{2}}{g},
\phi\right).
\end{eqnarray}
In what follows we understand $N(\omega, \phi)$ according to (8).
    We introduce also angle-independent spectra
\begin{eqnarray}
&&N(\omega)=\int^{2\pi}_{0}N(\omega, \phi)\,d\phi ,\nonumber\\
&&E(\omega)=\omega \,\frac{N(\omega)}{2\pi},\nonumber\\
&&M_{x}(\omega)=\frac{\omega^{2}}{g}\int^{2\pi}_{0}N(\omega,
\phi)\cos(\phi)\,d\phi . \nonumber
\end{eqnarray}
The conservative quantities in new variables take form
\begin{eqnarray}
&&N=\int^{\infty}_{0} N(\omega)\, d\omega ,\nonumber\\
&&E=\int^{\infty}_{0} E(\omega)\, d\omega , \nonumber\\
&&M_{x}=\int^{\infty}_{0}M_{x}(\omega)\,d \omega , \nonumber
\end{eqnarray}
and conservation laws of these quantities can be written in the
differential form
\begin{eqnarray}
&&\frac{\partial N(\omega)}{\partial t}=\frac{\partial Q}{\partial
\omega},\nonumber\\
&&\frac{\partial E(\omega)}{\partial t}=-\frac{\partial P}{\partial
\omega},\nonumber\\
&&\frac{\partial M_{x}(\omega)}{\partial t}=-\frac{\partial K}{\partial
\omega}.\nonumber
\end{eqnarray}
 Here $Q$ is the flux of wave action to small wave numbers, while $P$
and $K$ are fluxes of energy and momentum directed to high wave numbers. A
constant of motion is "real" if the corresponding flux is zero both at zero and infinite frequencies.
 Otherwise it is just a "formal" motion constant [Zakharov, Pushkarev,
 2000].

  Now let us introduce the differential operator
$$
L=\frac{1}{2}\frac{\partial^{2}}{\partial
\omega^{2}}+\frac{1}{\omega^{2}}\frac{\partial^{2}}{\partial \phi^{2}}
$$
 and present kinetic equation (7) in the form
\begin{equation}
\frac{\partial N(\omega, \phi)}{\partial t}=LA .
\end{equation}
Here
$$
A(\omega, \phi)=L^{-1}\,S_{nl},
$$
 and  $A$ is a result of action on $N(\omega, \phi)$ of a nonlinear
integral operator
\begin{eqnarray}
A(\omega, \phi)&=&\int F(\omega, \omega_{1}, \omega_{2}, \omega_{3},
\phi-\phi_{1},\phi-\phi_{2},\phi-\phi_{3})\nonumber\\
&&\times N(\omega_{1},\phi_{1})\,N(\omega_{2},\phi_{2})\,N(\omega_{3},\phi_{3})\nonumber\\
&&\hspace{0.5in}\times d\omega_{1}\, d\omega_{2}\, d\omega_{3}\, d\phi_{1}\, d\phi_{2}\, d\phi_{3}
.
\end{eqnarray}

    The explicit expression for $F$ is given in Appendix. $F$ is a
homogeneous function of order 12:
$$
F(\epsilon \omega, \epsilon \omega_{1},\epsilon \omega_{2}, \epsilon \omega_{3})=\epsilon^{12}F(\omega, \omega_{1}, \omega_{2},
\omega_{3})\sim g^{-4}\omega^{12} .
$$

 Further, if we denote
\begin{eqnarray}
&&A(\omega)=\frac{1}{2\pi}\int^{2\pi}_{0}A(\omega, \phi)\,d\phi ,\nonumber\\
&&B(\omega)=\frac{1}{2\pi}\int^{2\pi}_{0}B(\omega, \phi)\cos\phi \, d\phi ,\nonumber
\end{eqnarray}
then the fluxes $Q$, $P$, $K$ can be expressed in terms of $A$, $B$ in the
following form
\begin{eqnarray}
&&Q=\frac{\partial A}{\partial \omega},\\
&&P=A-\omega\,\frac{\partial A}{\partial \omega},\\
&&K=\frac{\omega}{g}\left(2B-\omega\frac{\partial B}{\partial \omega}\right).
\end{eqnarray}

    Formulae (11-13) are of key importance for the theory of
weak-turbulent spectra.

\section{Kolmogorov spectra}

    In this chapter we study solutions of the stationary equation
\begin{equation}
S_{nl}=0 ,
\end{equation}
 which is equivalent to equation
\begin{equation}
LA=0 .
\end{equation}
    One class of solutions for (15) is given by solution of the
equation
$$
A=0 ,
$$
and if these solutions exist, they are thermodynamic Rayley-Jeans
spectra
$$
n_{\vec k}=\frac{T}{\omega_{\vec k}+\mu}.
$$
In the case of surface gravity waves these solutions do not exist
because of the divergence of integrals in the operator $A$.
To get physically significant solutions we can partially integrate
equation (15) and put
\begin{equation}
A(\omega, \phi)=\omega Q+P+\frac{2Kg\cos\phi}{\omega} .
\end{equation}
In this case,
\begin{eqnarray}
&&A(\omega)=\omega Q+P,\nonumber\\
&&B(\omega)=\frac{Kg}{\omega}.
\end{eqnarray}
 By substituting (17) into (11-13) we see that constants
$Q$, $P$, and $K$ in both cases are the same.

    Equation (16) defines the most general Kolmogorov-type solution of
stationary kinetic equation; due to homogeneity of operator $A$ this equation
can be written in the form
\begin{equation}
N(\omega, \phi)=\frac{g^{\frac{4}{3}}P^{\frac{1}{3}}}{\omega^{5}}R\left(\frac{\omega
Q}{P}, \frac{2kg}{\omega P}, \phi\right)
\end{equation}
with the energy spectrum
\begin{equation}
E(\omega, \phi)=\frac{g^{\frac{4}{3}}P^{\frac{1}{3}}}{\omega^{4}}R\left(\frac{\omega
Q}{P}, \frac{2kg}{\omega P}, \phi\right) .
\end{equation}

    Let us study the most important special cases. If $Q=0$, $K=0$,
formulae (18), (19) give the Zakharov-Filonenko Kolmogorov spectrum of the
direct cascade
\begin{eqnarray}
N(\omega, \phi)&=&\frac{C_{p\,}g^{\frac{4}{3}}\,P^{\frac{1}{3}}}{\omega^5},\nonumber\\
E_{\omega}&=&\frac{C_{p}\,g^{\frac{4}{3}}\,P^{\frac{1}{3}}}{\omega^4}.
\end{eqnarray}
Here
$$
C_{p}=R(0,0,0)
$$
is the Kolmogorov constant of direct cascade (first Kolmogorov
constant). We can offer another definition of $C_{p}$.

 Suppose that $N(\omega,\phi)$ is an isotropic powerlike function of
    $\omega$,
\begin{equation}
N(\omega)=\omega^{-x}.
\end{equation}
 Special consideration (which is not at home in this article) shows
that integrals in $A$ converge if
$$
0<x<\frac{19}{4}.
$$
 Plugging (21) to (10) we obtain
\begin{equation}
A(\omega)=f(x)\,\omega^{(15-3x)} .
\end{equation}
Apparently,
$$
\left.f\right|_{x=5}=\frac{1}{C_{p}^{3}} .
$$

    Spectrum (20) has a clear physical interpretation; this spectrum is a direct
analog of the classical Kolmogorov spectrum of turbulence in an
incompressible fluid. This spectrum is realized if there is a source of energy at small wave
numbers and a sink of energy at high frequency region.

    The most general isotropic solution appears if $K=0$; then the spectrum
is
\begin{equation}
E_{\omega}=\frac{g^{\frac{4}{3}}P^{\frac{1}{3}}}{\omega^{4}}F\left(\frac{\omega
Q}{P}\right) .
\end{equation}
 Function  $F(\xi)$ depends on one variable; obviously $F(0)=C_{p}$. If
$\xi\to\infty$, spectrum should not depend on $P$.
 Hence $F(\xi)\to C_{q}/C_{p}\,\xi^{1/3}$ as
$\xi\rightarrow\infty$, equation (19) goes to Zakharov-Zaslavskii spectrum of
inverse cascade
\begin{equation}
E(\omega)=\frac{g^{\frac{4}{3}}C_{q}Q^{\frac{1}{3}}}{\omega^{\frac{11}{3}}}
,
\end{equation}
where $C_{q}$ is the Kolmogorov constant of the inverse cascade. Spectrum
(23) presumes that there is a source of wave action Q at high
frequencies and sink of wave action at small frequencies.

    A general isotropic spectrum (23) is realized if there exists
simultaneously a source of energy and sink of wave action at small
frequencies together with energy sink and wave action source
at high frequencies.

    If we study the most anisotropic case $Q=0$, $P=0$, then equation
(16) has the following solution
\begin{equation}
N(\omega,
\phi)=\frac{g^{\frac{4}{3}}\,h(\phi)\,(Kg)^{\frac{1}{3}}}{\omega^{\frac{13}{3}}}.
\end{equation}
From the symmetry consideration we can make a conclusion that
$$
h(\phi)=-h\,(\pi-\phi).
$$
Hence solution (25) is not positive at all values. This is a reason
to doubt that the general solution (19) is essentially positive and can
be realized in the whole $(\omega,\phi)$ plane. Anyway, it can be used for
approximation of real spectra in some finite part of wave-vector plane.

    In the important case of $Q=0$, solution (19) takes the form
\begin{equation}
E(\omega,\phi)=\frac{g^{\frac{4}{3}}P^{\frac{1}{3}}}{\omega^4}H\left(\frac{gK}{\omega
P},\phi\right).
\end{equation}

    Spectrum (26) can be found at small values of $g K/\omega P$.
Expanding in the Taylor series on this parameter, we obtain
$$
E(\omega,\phi)=\frac{g^{\frac{4}{3}}\,P^{\frac{1}{3}}}{\omega^4}\left(C_{p}+\frac{\alpha(\phi)\,
gK}{\omega P}+ \cdots\right) .
$$
 The correction to the isotropic spectrum satisfies the linearized
equation (16). As far as this situation is invariant with respect to
rotations in the $(\omega,\phi)$ plane,
$$
\alpha(\phi)=C_{2}\cos\phi .
$$
 Coefficient $C_{2}$ is known as the second Kolmogorov constant.

    If $\omega\rightarrow\infty$, then the right hand in (16) becomes
independent of angle. This means that the Kolmogorov solution
becomes isotropic at large $\omega$. The real observed spectra remain anisotropic
for arbitrary large $\omega$. The explanation is the following: in real
situation the momentum flux $K$ is not constant but is approximately
proportional to frequency.

\section{Local diffusion approximation}

    Many important features of the weak-turbulent theory can be understood in
a framework of a very simple theory.

    Let us accept the following approximation for the operator $A$ [Pushkarev, Zakharov, 1999]:
\begin{equation}
A(\omega,\phi)=\frac{a\,\omega^{15}}{g^{4}}\,N^{3}.
\end{equation}
 Here $a$ is a dimensionless constant, which should be found by comparison with experiment.
     In this case the kinetic
equation turns to the nonlinear diffusion equation
\begin{equation}
\frac{\partial N}{\partial t}=\frac{a}{g^4} \left(\frac{1}{2}\,\frac{\partial^2}
{\partial\omega^2}+\frac{1}{\omega^2}\,\frac{\partial^2}{\partial\phi^2}\right)\omega^{15}\,N^{3}.
\end{equation}

    Function $A$ has a very simple form and a general Kolmogorov solution is
\begin{equation}
E(\omega,\phi)=\frac{g^{\frac{4}{3}}\,P^{\frac{1}{3}}}{a^{\frac{1}{3}}\,\omega^{4}}\left(1+\frac{\omega\,
Q}{P}+\frac{2K\,g\cos\phi}{\omega P}\right)^{\frac{1}{3}}.
\end{equation}
 Now
\begin{eqnarray}
&&C_{p}=C_{q}=a^{-\frac{1}{3}},\nonumber\\
&&h(\phi)=\left(\frac{2\cos\phi}{a}\right)^{\frac{1}{3}}.\nonumber
\end{eqnarray}

Solution (29) is positive for all angles but in the case of large enough
frequencies only; that satisfies the inequality
$$
\frac{\omega P}{2kg}\left(1+\frac{\omega Q}{P}\right)>1 .
$$
Comparing (26) with (9) we find that in this case
\begin{eqnarray}
A(\omega, \phi)&=& \frac{a}{g^4}\,\omega^{15}\,N^3,\nonumber\\
A(\omega)&=&\frac{a}{g^4}\,\omega^{15}\,\frac{1}{2\pi}\int_{0}^{2\pi}N^3\,d\phi,\nonumber\\
B(\omega)&=&\frac{a}{g^4}\,\omega^{15}\,\frac{1}{2\pi}\int_{0}^{2\pi}n^3\cos\phi\,d\phi,\nonumber
\end{eqnarray}
and for the general Kolmogorov solution we obtain
  \begin{eqnarray}
  A(\omega,\phi)&=&P+\omega\,Q+\frac{2Kg\cos\phi}{\omega},\nonumber\\
  A(\omega)&=&P+\omega\,Q,\nonumber\\
  B(\omega)&=&\frac{K g}{\omega}. \nonumber
\end{eqnarray}

Both $A(\omega), B(\omega)$ are essentially positive. This is correct for
the general nonlocal case (12).
The formula for $A(\omega,\phi)$ presumes that a solution has sources of energy and momentum,
$P$ and $K$, as well as a sink of wave action $Q$ at the point $\omega=0$.

In a framework of the local approximation we can efficiently study a forced
stationary equation
\begin{equation}
S_{nl}+S_{in}+S_{ds}=0.
\end{equation}
We can assume that
$$
S_{in}+S_{ds}=\beta(\omega,\theta)\,N(\omega,\theta),
$$
and restrict our consideration by the case of angular symmetry
$\beta=\beta(\omega)$ only. Then equation (30) reads
\begin{equation}
\frac{a}{2g^4}\,\frac{\partial^2}{\partial\omega^2}\,\omega^{15}\,N^3 +\beta(\omega)\,N=0.
\end{equation}
Another form of this equation is the following:
\begin{equation}
\frac{\partial^2}{\partial\omega^2}A+\frac{g^{4/3}}{a^{1/3}}\,\frac{\beta(\omega)}{\omega^5}\,A^{1/3}=0.
\end{equation}
In a real situation solution $N(\omega)$ is concentrated in a finite
frequency band
\begin{eqnarray}
&&N>0, \quad \omega_1<\omega<\omega_2 ,\nonumber\\
&&N=0, \quad\omega<\omega_1 ,\,\,\omega>\omega_2 .
\end{eqnarray}
From continuity of $N$ and $\partial N/\partial\omega$ we obtain
\begin{equation}
N\left.\right|_{\omega=\omega_1}=0,\,\,\, \left. \frac{\partial N}{\partial\omega}
\right |_{\omega=\omega_1}=0,\,\,\,
N|_{\omega=\omega_2}=0,\,\,\,\left.\frac{\partial N}{\partial\omega}\right |_{\omega=\omega_2}=0.
\end{equation}
Conditions (33) define the boundary value problem for equations
(31), (32).

Since in neighborhood of the ends of interval (33) there exists asymptotics
\begin{eqnarray}
A&\simeq &\frac{1}{6} P_1 (\omega -\omega_1)^3 \quad P_1>0\nonumber\\
A&=&\frac{1}{6} P_2(\omega_2 -\omega)^3 \quad P_2>0
\end{eqnarray}
we obtain from (32) the following expressions for $\beta$:
\begin{eqnarray}
&&\beta(\omega_1)=-P_1^2\,\omega_1^5\,\left(\frac{6a}{g^4}\right)^{1/3},\nonumber\\
&&\beta(\omega_2)=-P_2^2\,\omega_2^5\left(\frac{6a}{g^4}\right)^{1/3}.
\end{eqnarray}
We can see now that a boundary problem has nontrivial solutions only if
$\beta(\omega)$ is negative at both ends of interval $\omega_1<\omega<\omega_2$. This
conclusion is very general. To get a stationary solution of equation
(30) we should have sinks both in low and high frequency regions. This
statement without a proof can be found in the paper [Komen, Hasselmann,
Hasselmann, 1984].

Condition (34) impose four restrictions on function $N(\omega)$ satisfying
to a second order ODE. This is not too much because the ends of the interval
$\omega_1<\omega<\omega_2$ are unknown. They can be found from the following conditions
for wave action and energy balance:
\begin{eqnarray}
&&\int_{\omega_1}^{\omega_2}\beta(\omega)\,N(\omega)\,d\omega=0,\nonumber\\
&&\int_{\omega_1}^{\omega_2}\omega\,\beta(\omega)\,N(\omega)\,d\omega=0.\nonumber
\end{eqnarray}
To satisfy the balance condition, we should have at least one domain of
instability, where $\beta(\omega)>0$, inside the interval $\omega_1<\omega<\omega_2$.
In a typical situation there is one such area. In this case $A(\omega)$ has
only one maximum at a point $\omega_3$ ($\omega_1<\omega_3<\omega_2$), and the whole
interval could be divided into three domains:

1. Area, where $A(\omega)$ grows. Suppose, some interval $A(\omega)$  can be
approximated by a linear function
$$
A(\omega)=Q(\omega -\omega_0).
$$
In this area,
\begin{eqnarray}
&&Q=\frac{\partial A}{\partial\omega}=const,\nonumber\\
&&P=A-\omega\frac{\partial A}{\partial\omega}\sim -\omega_0\,Q<0 .\nonumber
\end{eqnarray}
This is the area of \underline{inverse cascade}. A margin of this area is a frequency
$\omega^*$, where flux of energy $P$ changes the sign $P(\omega^*)=0$.

2. Area near $\omega\simeq\omega^*$, where $A(\omega)$ is almost constant. Here
$Q$ is small, while $P=A(\omega^*)$ is large and positive. This is the area
of \underline{direct cascade}.

3. Area of dissipation, where $\partial A/\partial\omega<0$. In this area $P>0,\,Q<0$.
Both the energy and the wave action are carried out to a zone of high
frequencies.

In the area of direct cascade, equation (32) can be approximately
integrated. We can rewrite this equation
\begin{equation}
\frac{\partial}{\partial\omega}\left(\omega\,\frac{\partial
A}{\partial\omega}-A\right)+\frac{g^{4/3}}{A^{1/3}}\,\frac{\beta(\omega)}{\omega^4}\,A^{1/3}=0,
\end{equation}
and put
$$
\omega\,\frac{\partial A}{\partial\omega}\ll A.
$$
This makes possible to integrate (35); the integration yields
\begin{eqnarray}
&&\frac{\partial
A}{\partial\omega}=\frac{g^{4/3}}{\omega^{1/4}}\,\frac{\beta(\omega)}{\omega^4}\,A^{1/3},\nonumber\\
&&A^{2/3}=\frac{2}{3}\,\frac{g^{4/3}}{a^{1/3}}\,\int_{\omega_0^*}^{\omega}\frac{\beta(\omega)}{\omega^4}\,d\omega
.\nonumber
\end{eqnarray}
In this approximation
\begin{eqnarray}
&&P=A+P_0\,\ln\left(\frac{\omega}{\omega^*}\right)^{3/2},\nonumber\\
&&P_0=\frac{g^2}{a^{1/2}}\left(\frac{2}{3}c\right)^{3/2},\nonumber
\end{eqnarray}
and $P_0$ is a slow function of $\omega$. For the
spectrum in the direct cascade area we have
\begin{equation}
E(\omega)\simeq P_0^{1/3}\frac{\left(\ln \frac{\omega}{\omega^*}\right)^{1/2}}{\omega^4}.
\end{equation}
Since in experiments $\omega^* \leq \omega_p$ ($\omega_p$ is a
frequency of spectral peak), at the current level of experimental
accuracy it is not easy to distinguish formula (36) from ZF spectrum
$\omega^{-4}$.

We should stress again that the forced stationary equation (28) has a
regular solution if and only if there are regions of intensive  damping
both in small and high wave numbers. What happens in other cases? Suppose,
there is no damping at all. In other words,
\begin{eqnarray}
&&\beta(\omega)>0,\quad \omega_1<\omega<\omega_2 ,\nonumber\\
&&\beta(\omega)=0, \quad \omega<\omega_1 .
,\,\,\,\,\omega>\omega_2\nonumber
\end{eqnarray}
Then in the isotropic case, outside of the forcing areas the spectra turn
to Kolmogorov type spectra
\begin{eqnarray}
&&\epsilon(\omega)= \frac{g^{4/3}\,P^{1/3}}{a^{1/3}\,\omega^4}, \quad
\omega>\omega_2 ,\nonumber\\
&&\epsilon(\omega)= \frac{g^{4/3}\,Q^{1/3}}{a^{1/3}\,\omega^{11/3}}, \quad
\omega<\omega_1 ,
\end{eqnarray}
and the fluxes of energy and wave action have the form
\begin{eqnarray}
&&P=\int_{\omega_1}^{\omega_2} \beta(\omega)\,\omega
\,\epsilon(\omega)\,d\omega ,\nonumber\\
&&Q=\int_{\omega_1}^{\omega_2} \beta(\omega)
\,\epsilon(\omega)\,d\omega .
\end{eqnarray}

There is a difference of principle importance between the area of direct
cascade $\omega>\omega_2$ and the area of inverse cascade
$\omega<\omega_1$. In the area of direct cascade, the integrals of motion,
energy and wave action are finite. On the opposite, in the inverse cascade
area the energy and the wave action diverge. We can say that the direct cascade has
finite capacity, while inverse cascade has infinite capacity.

A situation with no sink at high wave numbers is pure theoretical. Such a
sink always exists due to pletora of physical reasons: viscosity,
transformation of gravity waves to capillary waves, and finally due to
wave breaking.  A scrupulous consideration of these processes is not
necessary for understanding: what happens near the spectral peak?
Moreover, according to our preliminary study, Kolmogorov spectrum of direct
cascade can be formed in a finite time [Pushkarev, Resio, Zakharov, 2000].

On the contrary, the Kolnogorov spectrum of inverse cascade, due to its
infinite capacity, cannot be formed in a finite time. If there is no
intensive enough damping at small wave numbers, the inverse cascade cannot
be arrested. The downshift of spectral peak to small frequency area will
continue infinitely until it will be stopped by topographical (better to
say, geographical) factors.

\section{Discussion}

The first weak-turbulent Kolmogorov spectrum for gravity waves,
$\epsilon_{\omega}\simeq \omega^{-4}$, was derived analytically as an
exact solution of the kinetic equation in 1966 [Zakharov, Filonenko, 1966]. The
second Kolmogorov spectrum $\epsilon_{\omega}\simeq \omega^{-11/3}$ was
obtained in the same year in my PhD thesis in Novosibirsk [Zakharov,
1966]; in a regular journal the spectrum was published in 1982 [Zakharov,
Zaslavskii, 1982]. It is called now Zakharov-Zaslavskii (ZZ) spectrum.

For the first time, the spectrum $\epsilon_{\omega}\simeq \omega^{-4}$ was
observed experimentally by Toba in 1972. Since that this spectrum was
observed by many researches [Forrestal, 1981; Kahma, 1981; Kawai
et al, 1977; Donelan et al, 1985]. In 1987 Battjes et al found that
spectrum $\omega^{-4}$ fits the JONSWAP experiment much better than
$\omega^{-5}$. In 1985 O. Phillips published a well known article, where
he admitted that $\omega^{-4}$ spectrum fits the experiment better than
the "Phillips spectrum" $\omega^{-5}$. However he did not offer a proper
theoretical explanation of this fact.

In 1982-83 Zakharov and Zaslavskii published in the Russian journal four articles on application
of the weak turbulence theory to the wind-driven sea [Zakharov, Zaslavskii. 1982; 1983]; soon after S.
Kitaigorodskii used successfully these results for interpretation of experimental
data [Kitaigorodskii, 1983]. Since that time the weak turbulent theory became known to the
world community of oceanographers; however even now this theory is not completely
accepted. For almost thirty years the obvious facts:

1. $\epsilon_{\omega}\sim \omega^{-4}$ is the exact solution of the
stationary kinetic equation,

2. $\epsilon_{\omega}\sim \omega^{-4}$ is the spectrum persistently
observed in all experiments,

coexist separately in the collective conscience, almost not interacting
with each other.  I believe, this is a unique situation in the history of
science.

The standard arguments against the weak-turbulent theory are the
following (see, for instance, Komen, Cavaliery et al, 1994).

1. Zakharov-Filonenko spectrum is isotropic, while the real spectra are
anisotropic.

This argument is not a serious one. The isotropic spectrum $\omega^{-4}$
is the simplest example of weak-turbulent spectra. We showed in this paper
that more general Kolmogorov spectra, which carry momentum to high
frequency region, are anisotropic. Anyway, they are very close to
$\omega^{-4}$.

2. In the "classical" theory of turbulence for incompressible fluid, a
source of energy is concentrated in small wave numbers while in the case
of gravity waves the source of energy is distributed along the whole
spectrum.

One can add that the source of momentum is concentrated mostly in short
waves. The answer is the following: in a real situation the flux of energy is not a constant inside the
universal interval; this flux is a slowly growing function of frequency.
This leads to nonessential modification: the appearance of slowly growing
pre-factor proportional to $(\ln \omega/\omega^*)^{1/2}$.

We should stress again that the forced stationary kinetic equation (30)
has a solution that contains a finite amount of energy if and only if
there is an intensive dissipation in the low frequency region. This
dissipation arrests the inverse cascade and is essential in the spectral
peak area. Thus in this area equation (30) can be reduced to the form
$$
S_{nl}+S_{ds}=0.
$$
We like to stress that the physical origin of this low frequency
dissipation is unclear; the very fact of existence of this dissipation and the whole concept of
the "full-developed" are "mature" sea is
questionable. We will discuss this subject in details in another article.

In the universal region behind the spectral peak, the solutions of the
full equation (30) can be treated as solutions of the simple stationary
equation
$$
S_{nl}=0
$$
with frequency-dependent values of energy, wave action, and momentum
flux. This is the central point of the theory of weak turbulence.  We can
add that this point is supported now by massive numerical experiments (see
for instance  [Badulin et al, 2002], [Lavrenov et al, 2002], [Pushkarev et
al, to be published]).

\medskip

The research presented in this paper was supported by NSF grant NDM0072803
and by the Army Corps of Engineers, RTDIE  program, grant DACA
42-00-C0044.

\section*{Appendix}

Solution of the equation
\begin{equation}
LA=\left(\frac{1}{2}\,\frac{\partial^2}{\partial
\omega^2}+\frac{1}{\omega^2}\,\frac{\partial^2}{\partial
\phi^2}\right)\,A=f(\omega,\,\phi)
\end{equation}
is given by the Green function
$$
A(\omega,\,\phi)=\int_0^{\infty}\int_0^{2\pi}\,G(\omega,\,\omega',\,\phi
-\phi')\,f(\omega',\,\phi')\,d\omega'\,d\phi',
$$
where
\begin{eqnarray}
&&G(\omega,\,\omega',\,\phi-\phi')=-\frac{1}{2\pi}\sum_{n=-\infty}^{\infty}
\frac{\sqrt{\omega\,\omega'}}{\Delta_n}
\,e^{in(\phi
-\phi')} \nonumber\\
&&\times\left[\left(\frac{\omega'}{\omega}\right)^{\Delta_n}\,\Theta\left(1-\frac{\omega'}
{\omega}\right)+
\left(\frac{\omega}{\omega'}\right)^{\Delta_n}\,\Theta \left(\frac{\omega'}
{\omega}-1\right)\right]\nonumber
\end{eqnarray}
Here
$$
\Delta_n=\sqrt{\frac{1}{4}+2n^2}
$$
and
$$\Theta(\xi)=\left\{
\begin{array}{cc}
1&\xi>0\\
0&\xi<0
\end{array}
\right\}
$$
Equation (10) appears after substituting of $S_{nl}(\omega',\,\phi')$ as
$f(\omega',\,\phi')$ in formula (40).

\end{document}